\documentclass{aa}
\usepackage{graphicx}
\usepackage{epsfig}
\usepackage{txfonts}
\def\deg       {$^{\circ}$}

\def \sig      {$\sigma$}
\def \gray     {$\gamma$-ray}
\def \grays    {$\gamma$-rays}

\newcommand{\be}{\begin{equation}}
\newcommand{\ee}{\end{equation}}

\newcommand{\swift}{{\it Swift}}

\begin{document}
   \title{The multifrequency campaign on 3C 279 in January 2006}

   \author{W.~Collmar \inst{1},
           M. B\"ottcher \inst{2},
           T.P. Krichbaum \inst{3},
           I. Agudo \inst{4,5,3},
           E. Bottacini \inst{1},
           M. Bremer \inst{6},
           V. Burwitz \inst{1},
           A. Cuccchiara \inst{7},
           D. Grupe \inst{7},
           M. Gurwell \inst{8}
             }

   \offprints{W. Collmar}
   \mail{wec@mpe.mpg.de}

   \institute{
       Max-Planck-Institut f\"ur extraterrestrische Physik,
       Giessenbachstrasse, D-85748 Garching
     \and
       Astrophysical Institute, Department of Physics and Astronomy,
       Ohio University, Athens, OH 45701, USA
     \and
       Max-Planck-Institut f\"ur Radioastronomie, Auf dem H\"ugel 69,
       D-53121 Bonn, Germany
     \and
       Institute for Astrophysical Research, Boston University, 725 Commonwealth Avenue, Boston, MA 02215, USA
     \and
      Instituto de Astrof\'{\i}sica de Andaluc\'{\i}a (CSIC), Apartado 3004, E-18080 Granada, Spain
     \and
      IRAM, Avenida Divina Pastora 7, Local 20, E-18012 Granada, Spain
     \and
      Pennsylvania State University, 525 Davey Lab, University Park, PA 16802, USA
     \and 
      Harvard-Smithsonian Center for Astrophysics, 60 Garden Street, Cambridge, MA 02138, USA
   }

   \date{Received 20 April 2010 / Accepted 12 July 2010}

  \abstract
   { The prominent blazar 3C~279 is known for its large-amplitude variability 
     throughout the electromagnetic
     spectrum and its often $\gamma$-ray-dominated spectral energy distribution.
     However, the characterization of its broadband spectral variability still
     lacks a consistent picture, and the origin of its high-energy emission is
     still unclear.
   }
   { We intend to characterize the spectral energy distribution and spectral
     variability of 3C~279 in its optical high state.
   }
   { Prompted by an optical high state of 3C~279, we organized an extensive 
     multiwavelength campaign with coverage from radio to hard X-ray energies. 
     The core components of the campaign were INTEGRAL and Chandra ToO observations 
     in January 2006, augmented by X-ray data from Swift and RXTE as well as 
     radio through optical coverage. 
     }
   { The blazar was observed at a moderately high optical state. A well-covered
     multifrequency spectrum from radio to hard X-ray energies could be derived.
     During the flare, the radio spectrum was inverted, with a prominent
     spectral peak near 100 GHz, which propagated in time toward lower frequencies.
     The spectral energy distribution (SED) shows the typical two-bump shape,
     the signature of non-thermal emission from a relativistic jet. 
     As a result of the long exposure
     times of INTEGRAL and Chandra, the high-energy spectrum (0.3 -- 100~keV)
     was precisely measured, showing --- for the first time --- a possible
     downward curvature.
     A comparison of this SED from 2006 to the one observed in 2003,
     also centered on an INTEGRAL observation, but during an optical low-state,
     revealed the surprising fact that --- despite a significant change of the
     high-frequency synchrotron emission (near-IR/optical/UV) --- the low-energy
     end of the high-energy component (X-ray energies) remained virtually 
     unchanged compared to 2003.
     }
   { Our results prove that the two emission components do not vary simultaneously.
     This provides strong constraints on the modeling of the overall emission of
     3C~279. When interpreted with a steady-state leptonic model, the variability
     among the SEDs displaying almost identical X-ray spectra at low flux levels, 
     but drastically different IR/optical/UV fluxes, can be reproduced by a change 
     solely of the low-energy cutoff of the relativistic electron spectrum. 
     In an internal shock model for blazar emission, such a change could be 
     achieved through a varying relative Lorentz factor of colliding shells 
     producing internal shocks in the jet, and/or
     the efficiency of generating turbulent magnetic fields
     (e.g., through the Weibel instability) needed for efficient 
     energy transfer from protons to electrons behind the shock.
     }


   \authorrunning{W. Collmar et al.}
   \titlerunning{The Multifrequency Campaign on 3C~279 in January 2006} 
   \maketitle
%

\section{Introduction}
Blazars are the most variable and violent type of active galactic
nuclei (AGN). They show rapid variability in all wavelength bands,
polarized emission in the radio and optical bands, and often
superluminal motion. The blazar subclass of AGN combines
flat-spectrum radio quasars (FSRQs) and BL~Lac objects.
According to the unified model of AGN (e.g., \cite{Urry95}),
blazars are sources which expell jets close to our line of sight.
The discovery by the experiments aboard the Compton Gamma-Ray Observatory
(CGRO) that blazars can radiate a large --- sometimes even the major ---
fraction of their luminosity at \gray\ energies marked a milestone
in our knowledge of blazars. During the CGRO mission about 90 blazars
were detected by the different CGRO experiments at \gray\ energies, the
majority by the Energetic Gamma-Ray Experiment Telescope (EGRET) at
energies above $\sim$100~MeV (\cite{Hartman99}). It is generally accepted
that the radio to optical/UV continuum of blazars is synchrotron radiation
generated by relativistic electrons in a magnetized jet. For the high-energy
continuum from X- to \gray\ energies, two fundamentally different approaches
are being considered (for a recent review see, e.g., \cite{boettcher07}).
If protons are not accelerated to ultra-high energies, exceeding the 
threshold for photo-pion production on the co-spatially produced synchrotron 
radiation field, high-energy emission might be dominated by Comptonization 
of soft photons by the non-thermal jet electrons (leptonic models; e.g., 
\cite{maraschi92,ds93,sikora94}). Otherwise, the high-energy
radiative output might be dominated by proton-synchrotron radiation
as well as synchrotron and Compton emission from the decay products
of pions produced in photo-pion production processes (hadronic models;
e.g., \cite{mb92,aharonian00,muecke03}).

The optically violently variable (OVV) quasar 3C~279, located at a redshift
of 0.536, is one of the most prominent representatives of this source class.
The blazar shows rapid variability in all wavelength bands, polarized emission
in the radio and optical, superluminal motion, and a compact radio core with a
flat radio spectrum. The quasar 3C~279 was found to be a persistent --- though highly
variable --- emitter of $\gamma$-rays above 100~MeV by EGRET. Subsequently the
source was frequently observed simultaneously at radio and optical bands,
and was subject of many multiwavelength campaigns during the CGRO era
(e.g., \cite{Maraschi94,Hartman96,Wehrle98}). A complete compilation
of all simultaneous spectral energy distributions (SEDs) of 3C~279 collected
during the lifetime of CGRO, including modeling of those SEDs with a leptonic
jet model, is presented in \cite{Hartman01a}. The conclusion was that the
high-energy emission most likely consists --- especially during flaring epochs
--- of several emission components that vary independently. The synchrotron
self-Compton (SSC) mechanism (\cite{maraschi92,bm96}) dominates the
 X- and soft \gray\ part while at higher \gray\ energies an external Compton
component (e.g., \cite{ds93,sikora94}) becomes apparent.

Despite the enormous observational and theoretical multifrequency efforts,
the nature and origin of the high-energy (X-rays to \grays) emission of 3C~279
still remains unclear. Also, the connection of the high-energy radiation
to the low-energy radiation is not understood. Cross-correlations between
different energy bands, particularly between optical, X-ray, and \grays\,
were investigated by \cite{Hartman01b}, but no consistent trends could
be found. This poses a serious challenge to the quite often adopted single-zone
leptonic jet models, in which the same electron population produces both
the low-frequency emission through synchrotron radiation and the high-energy
emission through Compton scattering.

In order to improve on this general situation, \cite{Collmar04} continued
to observe 3C~279 in a multifrequency campaign organized around INTEGRAL 
(International Gamma-Ray Astrophysics Laboratory) high-energy observations. 
INTEGRAL was launched in October 2002. In a campaign in June 2003, 3C~279 
was detected significantly at hard X-ray energies for the first time by the 
INTEGRAL/IBIS experiment (\cite{Collmar04}). Those high-energy observations
were supplemented in X-rays by a short Chandra  pointing, and by ground based
monitoring from radio to optical bands. In June 2003, the blazar was found in
the faintest optical brightness (optical R-band: $\sim$17~mag) of the last 10
years, roughly 5~mag fainter than the maximum, and about 2.5 to 3 mag fainter
than average. That campaign allowed \cite{Collmar04} to compile a contemporaneous
spectral energy distribution (SED) of an exceptional optical low-state.

In order to measure emission changes as function of optical brightness,
we proposed for an intensive multifrequency campaign during an optical
high state including high-energy INTEGRAL observations. On 5 January 2006,
the trigger criterion was met when 3C 279 exceeded an R-band flux
corresponding to R $=$ 14.5~mag. INTEGRAL observations were carried out
together with supplementing multifrequency observations from radio via IR 
and optical to X-ray energies. Preliminary results from this campaign 
were reported by \cite{Collmar07}. Intensive ground-based monitoring
in the IR and optical was assured by a WEBT (Whole Earth Blazar Telescope)
campaign (see, e.g., \cite{Raiteri06,Villata07}, and references therein).
The results of the WEBT component of this campaign on 3C~279 in 2006 were 
presented in detail by \cite{bbj07}. In this paper we present details of the 
non-WEBT results of the radio, the X-ray, and soft \gray\ bands (Sect.
\ref{observations}), the interpretation of the radio data (Sect.~\ref{radioint}),
as well as the final SED (Sect.~\ref{SED}). In 
addition, we present a model interpretation of the SED, and discuss our 
results in the framework of leptonic emission scenarios (Sect.~\ref{modeling}).

Shortly after the core period of the multiwavelength campaign described here, 
the Major Atmospheric Gamma-Ray Imaging Cherenkov Telescope (MAGIC) reported 
the detection of very-high-energy (VHE: $\gtrsim 200$~GeV) $\gamma$-ray
emission from 3C~279 on 23 February 2006 (\cite{albert08}). This detection
was extremely surprising for two reasons: (1) FSRQs like 3C~279 are known
to have the peaks of their broad emission components in their SEDs at
relatively low energies. In particular, the $\gamma$-ray component of
3C~279 has been known from the EGRET era to peak at a few GeV, and the
$\gamma$-ray spectrum was generally not expected to extend out to VHE
frequencies at a substantial flux level. (2) 3C~279 is the VHE $\gamma$-ray
source with the highest confirmed redshift to date. At such large redshifts,
$\gamma\gamma$ absorption of VHE photons by the extragalactic background
light (e.g., \cite{dk05}) is expected to lead to severe attenuation of
the spectrum in the VHE regime. This prompted \cite{albert08} to propose
that the universe may be more transparent to VHE $\gamma$-rays than
previously thought (however, \cite{ss09} dispute this claim). It has 
been shown by \cite{boettcher09} that if this VHE detection of 3C~279 
can be confirmed, it would pose severe problems for any single-zone leptonic 
model interpretation and may potentially favor hadronic over leptonic models.

However, the reported VHE $\gamma$-ray flux, corrected for intergalactic
$\gamma\gamma$ absorption, lies substantially above a straight extrapolation of
non-simultaneous MeV -- GeV spectra measured by EGRET as well as more recently
by the Fermi Gamma-Ray Space Telescope (\cite{Abdo09}). Also, in spite of
intensive efforts, no other VHE $\gamma$-ray observatory has so far been
able to confirm the detection of VHE $\gamma$-ray emission from 3C~279. This
indicates that the detection of 23 February 2006 must correspond to an extraordinary
flaring activity and cannot be considered representative of the time of the
simultaneous multiwavelength observations reported here. Therefore, we will
focus our modeling efforts on leptonic models, which have been shown to provide
a successful description of all multiwavelength SEDs of 3C~279 throughout 
the CGRO era (\cite{Hartman01a}). 

For the purpose of the model interpretation, we will use the widely accepted 
$\Lambda$CDM cosmology with $H_0 = 70$~km~s$^{-1}$~Mpc$^{-1}$, $\Omega_m = 0.3$, 
and $\Omega_{\Lambda} = 0.7$. In this cosmology, the luminosity distance of 3C~279 
is $D_L = 3.08$~Gpc. Throughout the paper we refer to the energy spectral index 
$\alpha$, defined through $F_{\nu} [Jy] \propto \nu^{-\alpha}$, and the photon
spectral index $\Gamma_{\rm ph} = \alpha + 1$.


\section{\label{observations}Observations, data reduction, and results}

After 3C~279 met the trigger criterion for the targeted optical high-state ($R
\le 14.5$~mag) on 5 January 2006, we requested the pre-approved 
high-energy ToO observations with INTEGRAL and Chandra. To monitor the 
X-ray and UV-state of 3C~279 during the INTEGRAL 
observations, we proposed and were granted monitoring observations by Swift.
In addition, ground-based monitoring observations in the framework of a WEBT
campaign as well as sub-millimeter and radio monitoring of several telescopes
were contributed to our campaign. In this section, we describe details of the collection
and analysis of the previously unpublished data and briefly summarize the relevant
aspects of campaign contributions which were previously published elsewhere
(\cite{bbj07,boettcher09}). 

\subsection{\label{INTEGRAL}INTEGRAL}

INTEGRAL is an ESA scientific mission dedicated to high-resolution 
spectroscopy with SPI (\cite{vedrenne03}) and imaging with IBIS/ISGRI  
(angular resolution: $12'$ FWHM, point source location accuracy: $\simeq 1'-3'$; 
IBIS/ISGRI, see \cite{ubertini03,lebrun03}) of celestial sources in hard X- and soft 
$\gamma$-rays ($15$~keV to $10$~MeV). In addition, INTEGRAL 
provides simultaneous monitoring at X-rays ($3-35$~keV) with JEM-X (\cite{lund03})
and at optical wavelengths (Johnson V-filter) with OMC (\cite{mas03}). 

INTEGRAL observed the blazar in three pointings between 13 January and 22 January 
2006 for a total observation time of $\sim 512$~ks (Table~\ref{he_log}). Due to 
annealing, no INTEGRAL SPI data are available. The INTEGRAL analyses were done with 
the \emph{INTEGRAL} \texttt{Offline Scientific Analysis (OSA) version 7.0}.

\begin{figure}[th]
\centering
   \includegraphics[width=\columnwidth]{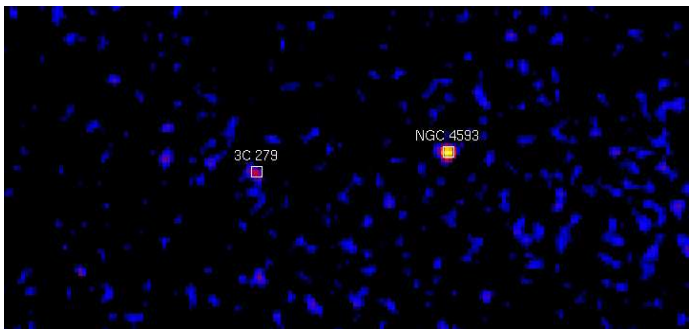}
\caption{
The ISGRI image shows a 6.4\sig\ detection of 3C~279 in the  
20-60~keV band. In addition, the Seyfert galaxy NGC 4593 is even 
more clearly detected. The image shown covers a sky region of 
$\sim$11.8\deg\,$\times$\,5.6\deg.
\label{ISGRI}
}
\end{figure}

The IBSI/ISGRI data were screened for data failures or problems, and in the end
a total observation time of 505,043~seconds, available in 149 so-called science 
windows, was selected for final analysis.
IBIS/ISGRI detected the blazar at energies between 20 and 100 keV with a significance 
of $\sim$7.5$\sigma$ during the total used exposure of 505~ks. The ISGRI image 
(Fig.~\ref{ISGRI}) of the 20-60~keV band shows a 6.4$\sigma$ detection of 3C~279. 
Other Virgo region sources are also detected in these observations, for example the 
prominent quasar 3C~273 and the Seyfert galaxy NGC~4593. The blazar was measured by 
ISGRI at a surprisingly low flux level of 
$(2.53 \pm 0.52) \times 10^{-4}$~ph~cm$^{-2}$~s$^{-1}$.  
The spectral analysis with XSPEC between 20 and 100 keV, assuming a power-law shape,
yields a weakly determined average spectral shape (Fig.~\ref{xspecfig}) of the photon 
index $\Gamma_{\rm ph} = 2.0 \pm 0.5$ (1$\sigma$).

Subdividing the observations into three roughly equal time intervals 
(13 --15 January, 16--18 January, and 18-20 January), 
yields a low significance flux value for each interval. 
No significant variability is seen, but there appears to be a trend of a rising 
hard X-ray flux (20 -- 60 keV) toward the end of INTEGRAL's observational period 
(Fig.~\ref{he_lc}).

JEM-X did not detect 3C~279. Upper limits were derived for two energy bands (5 -- 10 
and 10 -- 20 keV) based on the mosaic images from these observations. The OMC monitoring 
data were analyzed as well (see \cite{bottacini07}). We do not include them in this paper, 
because during the campaign superior optical measurements were performed by several 
ground-based observatories (see Sect.~\ref{WEBT}) as well as the Swift UVOT (Sect.~
\ref{Swift}). 

\begin{figure}[th]
\centering
   \includegraphics[width=\columnwidth]{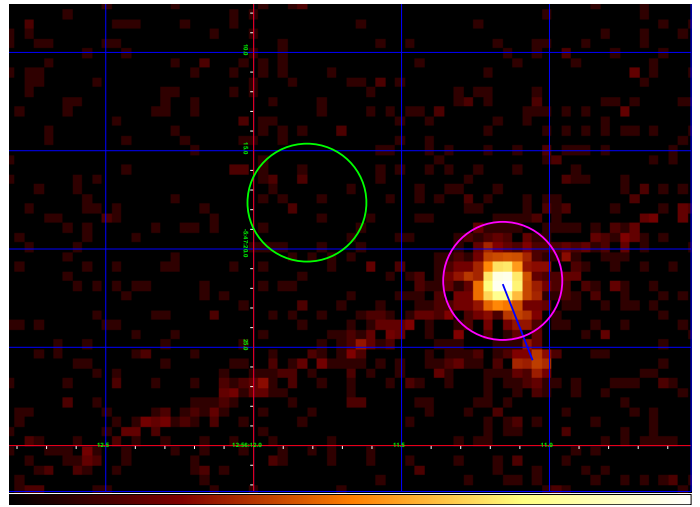}
\caption{
 The blazar is clearly visible in the Chandra image. A weaker source, 
probably the X-ray jet of 3C~279, is also visible at an angular distance 
of $\sim$4~arcsec. The purple (source) and the green (background) circles 
indicate the sky regions used for data analysis.
The image shown covers a sky region of $\sim$34.0\arcsec\,$\times$\,24.5\arcsec.
}
\label{chandrafig}
\end{figure}

\subsection{\label{Chandra}Chandra}

After the INTEGRAL observations were approved, we also triggered our pre-approved
Chandra 30~ks ToO observation on 3C~279. It was scheduled on 17 January 2006, 
roughly centered in time on the one week INTEGRAL observation. Chandra 
observed the blazar for a total observation time of $\sim$32~ks in order 
to determine contemporaneously and most accurately the X-ray state and
spectrum of 3C~279. To avoid pile-up in the Chandra detectors of the assumed 
strong X-ray source, the LETG-ACIS-S mode was used. 
 
\begin{figure}[th]
\centering
   \includegraphics[width=\columnwidth]{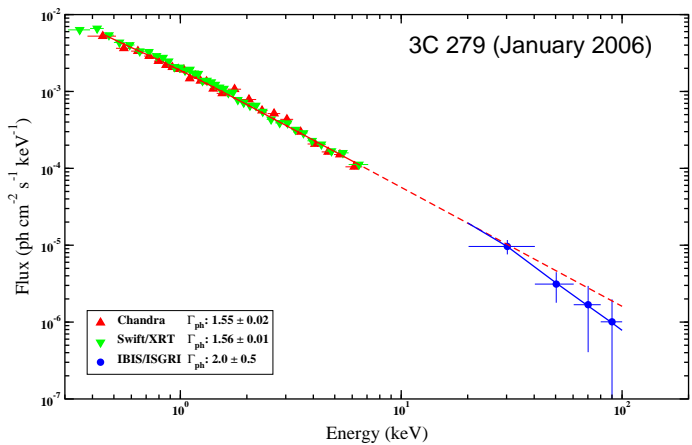}
\caption{
X- to hard X-ray spectrum of 3C~279 as measured by Chandra (17 January 2006; 
exposure $\sim$32~ks), Swift (6 pointings between 13 -- 19 January 2006; exposure
$\sim$65~ks) and INTEGRAL/ISGRI (3 pointings between 13 -- 22 January 2006; exposure
$\sim$505~ks) together with the Chandra (the Swift is completely consistent 
with Chandra) and ISGRI best-fit power-law shapes (solid lines). The dashed line 
shows the extrapolation of the Chandra spectrum up to 100~keV, indicating the softer 
ISGRI spectrum above $\sim$20~keV. The best-fitting photon indices including their 
errors (1$\sigma$) are given. 
}
\label{xspecfig}
\end{figure}

The blazar 3C~279 was significantly detected by Chandra. 
The Chandra X-ray image (Fig.~\ref{chandrafig}) shows next to it 
a weaker X-ray source at a distance of $\sim$4~arcsec. This could be emission 
from its jet because it is along the direction of its VLA jet (e.g. \cite{akujor94}). 
Chandra measured a well determined X-ray spectrum between 0.3 
and 7~keV (Fig.~\ref{xspecfig}).  Assuming the canonical power-law shape at
 X-ray energies, a spectral photon index of 
$\Gamma_{\rm ph} = 1.55 \pm 0.02$ was derived. 
The spectrum is consistent (reduced $\chi^{2}$: 1.14) with a hard power-law shape. 
However, the spectral analysis indicated a spectral bending from a harder 
to a softer spectrum toward higher energies. A fit with a broken powerlaw 
shape yielded a better fit (reduced $\chi^{2}$: 0.82): a harder shape (photon index: 
1.46$\pm$0.03) below $\sim$3~keV and softer one (photon index: 1.87$\pm$0.12) 
above. In particular, no soft excess was found, which would have indicated 
a contribution of the synchrotron component or the accretion disk to the 
soft X-ray emission. Chandra did not observe any variability of 
3C~279 during the uninterrupted source-pointing of $\sim$8 hours. 

\subsection {\label{Swift}\swift}

The $\gamma$-ray burst explorer \emph{Swift} was launched in 2004, and carries
three co-aligned detectors (\cite{gehrels04}): the Burst Alert Telescope 
(BAT, \cite{barthelmy05}), the X-Ray Telescope (XRT, \cite{burrows05})
and the Ultraviolet/Optical Telescope (UVOT, \cite{roming05}). Between 
13 and 19 January 2006, Swift regularly observed the blazar, and ---
by our ToO request --- contemporaneously with INTEGRAL (Table~\ref{he_log}). 
The data analysis followed the standard procedure for {\it Swift} XRT
data (e.g. \cite{grupe10}) by using the HEASOFT 6.8 tools and XSPEC 
12.5.1.n (\cite{arnaud96}). All observations were performed in photon-counting mode (\cite{hill04}). 
Source photons were selected in the circular region with a 
radius of 82.5$^{''}$. Spectral data were re-binned to 50 counts per
bin (except for segment 003 where we used 20 counts per bin).

\begin{figure}[ht]
\includegraphics[width=\columnwidth]{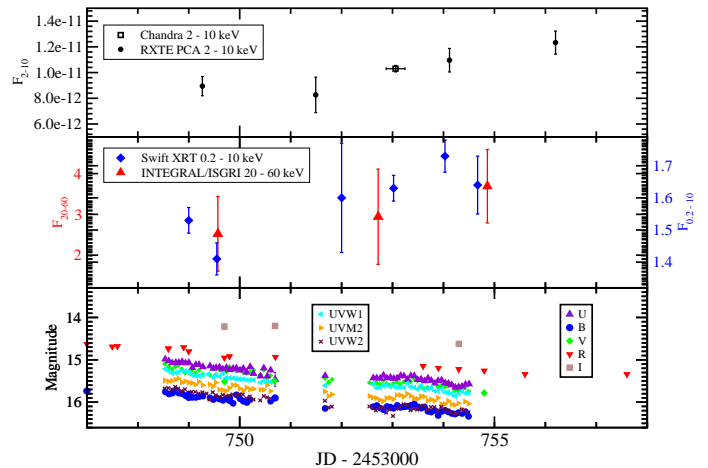}
\caption{X-ray (top two panels) and optical/UV (lower panel) light curves of 3C~279 
during the core campaign period, 13 -- 22 January 2006. Units for the X-ray light
curves are: $F_{2-10}$ [erg~cm$^{-2}$~s$^{-1}$] and 
$F_{20-60}$ [$10^{-4}$~ph~cm$^{-2}$~s$^{-1}$]}
\label{he_lc}
\end{figure}

The XRT measured a moderately variable 0.2 -- 10~keV X-ray flux with a 
variability amplitude of $\Delta F / F \lesssim 25$~\%. After an initial
decline throughout 13 January 2006, the Swift XRT flux seems to indicate a 
gradual rising trend during the remainder of the INTEGRAL observation period,
in agreement with the trend seen by RXTE in the 2 -- 10~keV range (see
Fig.~\ref{he_lc}). 

Parallel to XRT, UVOT monitored 3C~279's ultraviolet and optical emission.  
The UV magnitudes have been de-reddened using the extinction curves of
\cite{cardelli89} with $R_V = 3.1$ and $A_V = 0.095$, taken from the NASA/IPAC 
Extragalactic Database (NED\footnote{\tt http://nedwww.ipac.caltech.edu/}).
The data indicate a continuous fading of the source in all six 
filter bands (Figs.~\ref{he_lc} and \ref{mwlightcurve}), in agreement 
with the trend observed with ground-based optical observatories (see Sect.~\ref{WEBT}).

\begin{table*}[thb]
\caption[]{Observation log of the high-energy observations during our
January 2006 campaign. For INTEGRAL the orbital revolution number is given. 
}
\label{he_log}
\begin{flushleft}
\begin{tabular}{lcccr}
\hline\noalign{\smallskip}
Obs. Periods      &  T-Start  & T-Stop  & Exposure &  Satellite \\
\noalign{\smallskip}
\hline
\noalign{\smallskip}
001 & 2006-01-13 00:40 & 2006-01-13 23:37 & 15014 & Swift\\
002 & 2006-01-14 00:42 & 2006-01-15 02:35 & 10613 &      \\
003 & 2006-01-16 04:07 & 2006-01-16 23:34 &  1581 &      \\
004 & 2006-01-17 01:03 & 2006-01-17 23:53 & 19289 &      \\
005 & 2006-01-18 01:14 & 2006-01-18 24:00 & 13897 &      \\
006 & 2006-01-19 01:08 & 2006-01-19 06:14 &  4113 &      \\
\hline\noalign{\smallskip}
001 & 2006-01-17 08:59 & 2006-01-17 17:51 & 31920 & Chandra \\
\hline\noalign{\smallskip}
397 & 2006-01-13 00:22 & 2006-01-15 03:03 & 174879 & INTEGRAL \\
398/399 & 2006-01-16 07:17 & 2006-01-20 17:41 & 322326 &  \\
400 & 2006-01-22 05:13 & 2006-01-22 09:25 & 14392 &  \\
\noalign{\smallskip}
\hline
\end{tabular}\end{flushleft}
\begin{list}{}{}
\item[$^{\mathrm{a}}$] Start and end times are given in UT
\item[$^{\mathrm{b}}$] Exposure is given in s
\end{list}\end{table*}

\subsection{\label{RXTE}RXTE}

In the course of a long-term monitoring program led by A. Marscher 
of $\gamma$-ray bright 
blazars using the Proportional Counter Array (PCA) on board the Rossi 
X-Ray Timing Explorer (RXTE), 3C~279 was observed with 2 -- 3 pointings 
per week throughout (and beyond) the campaign period. Details of the 
observations and data analysis have been published in \cite{Chatterjee08}, 
and the X-ray data have been discussed in the context of the WEBT campaign 
(see Sect.~\ref{WEBT}) and the MAGIC detection in \cite{boettcher09}.
The RXTE data are included in Fig.~\ref{he_lc}, and the multiwavelength
light curves (radio through hard X-rays) throughout January and February 
2006 are shown in Fig.~\ref{mwlightcurve}. 

Before and during the core period of our multiwavelength campaign, the
RXTE light curve showed 3C~279 to be in a low flux state, near its historical
minimum, with a moderately soft spectrum with $\alpha \sim 1$ in the 2 -- 
10~keV range. Toward the end of January 2006, i.e., after the core campaign 
period, the object went into a high-activity X-ray state with average X-ray
flux a factor $\sim 2$ -- 3 higher than during the previous low state. In
this state, 3C~279 showed substantial, correlated flux and spectral variability 
on time scales of $\sim 10$~days. Notably, the time of the MAGIC detection 
occurred during the rising phase of a major X-ray outburst, peaking about one week
thereafter at a flux level of $\sim 5$ times the previous quiescent flux.
In the RXTE X-ray band 3C 279 is continuously variable 
with flux variations up to a factor of 10 ($F_{2-10}$ between 
$\sim$ 0.5 and 5 $\times$ 10$^{-11}$~erg~cm$^{-2}$~s$^{-1}$). 
See e.g. \cite{Marscher06} for a longterm (1996 - 2005) RXTE X-ray 
light curve of 3C 279.

\subsection{\label{WEBT}Optical monitoring}

\begin{figure}[th]
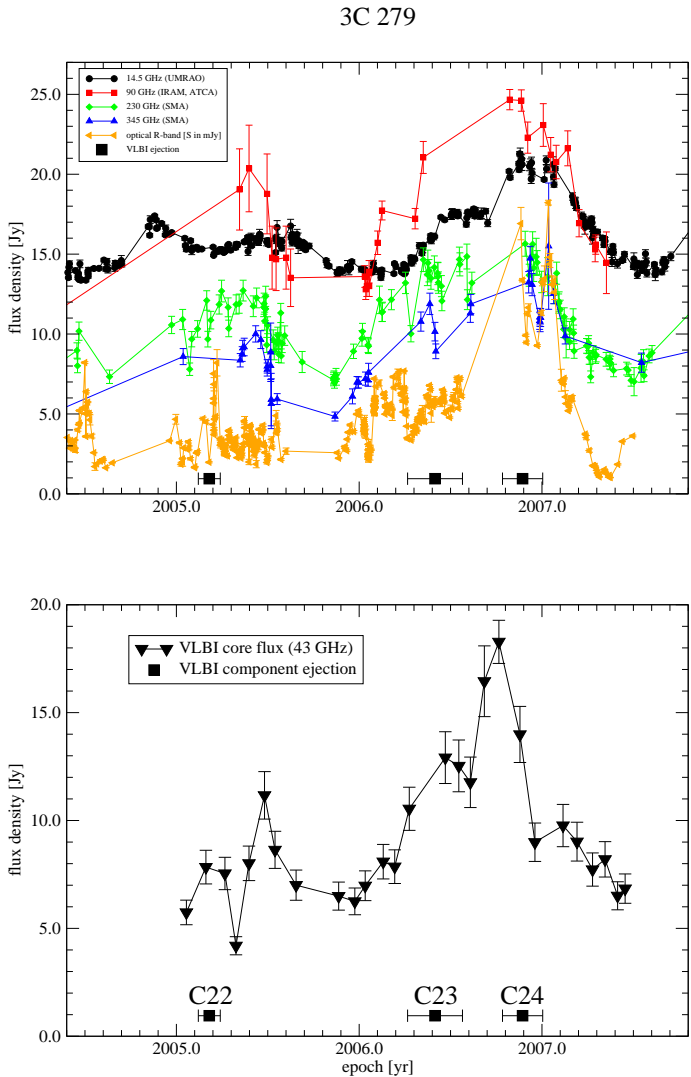

\vspace*{0.5cm}
\centering{
   \includegraphics[width=\columnwidth]{fig5a.eps} }
   ~~\\
centering{
   \includegraphics[width=\columnwidth]{fig5b.eps} }
\caption{Top: Radio and optical flux density plotted versus time. 
Different symbols and colors
denote  different observing frequencies: 14.5\,GHz black circles,
90\,GHz red squares, 230\,GHz green diamonds, 345\,GHz blue triangles, and 
optical R-band (amplitudes given in [mJy]) yellow triangles. Black squares
mark VLBI component ejection as indicated in the panel below.
Bottom: 43\,GHz VLBI core flux density versus time. Black squares with 
horizontal error bars denote the linearly back-extrapolated ejection times 
of the VLBI components C22, C23 and C24 (in order of their appearance). 
}
\label{radioplot1}
\end{figure}

\begin{figure}[th]
\centering
   \includegraphics[width=\columnwidth]{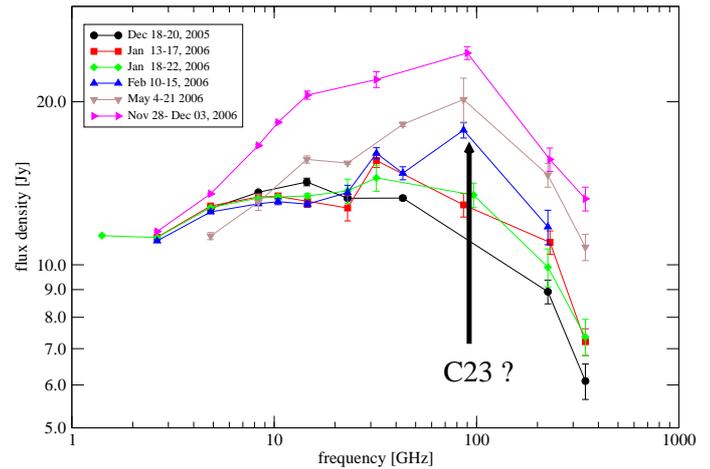}
\caption{Quasi-simultaneous broad-band radio spectra of 3C\,279. Different symbols
and colors denote different observing epochs, as indicated by the legend box
in the top left of the figure. Because of the irregular time sampling at
the different observatories, the binning intervals differ from epoch to epoch.
The spectrum of November 2006 was observed near the time of the peak of the 2007.0 radio flare,
see Fig.~\ref{radioplot1}. The solid black arrow marks the spectral flaring component,
which may be associated with the new VLBI jet components C23 in \cite{Larionov08}.
}
\label{radioplot2}
\end{figure}

\begin{figure*}[bt]
\includegraphics[width=17cm]{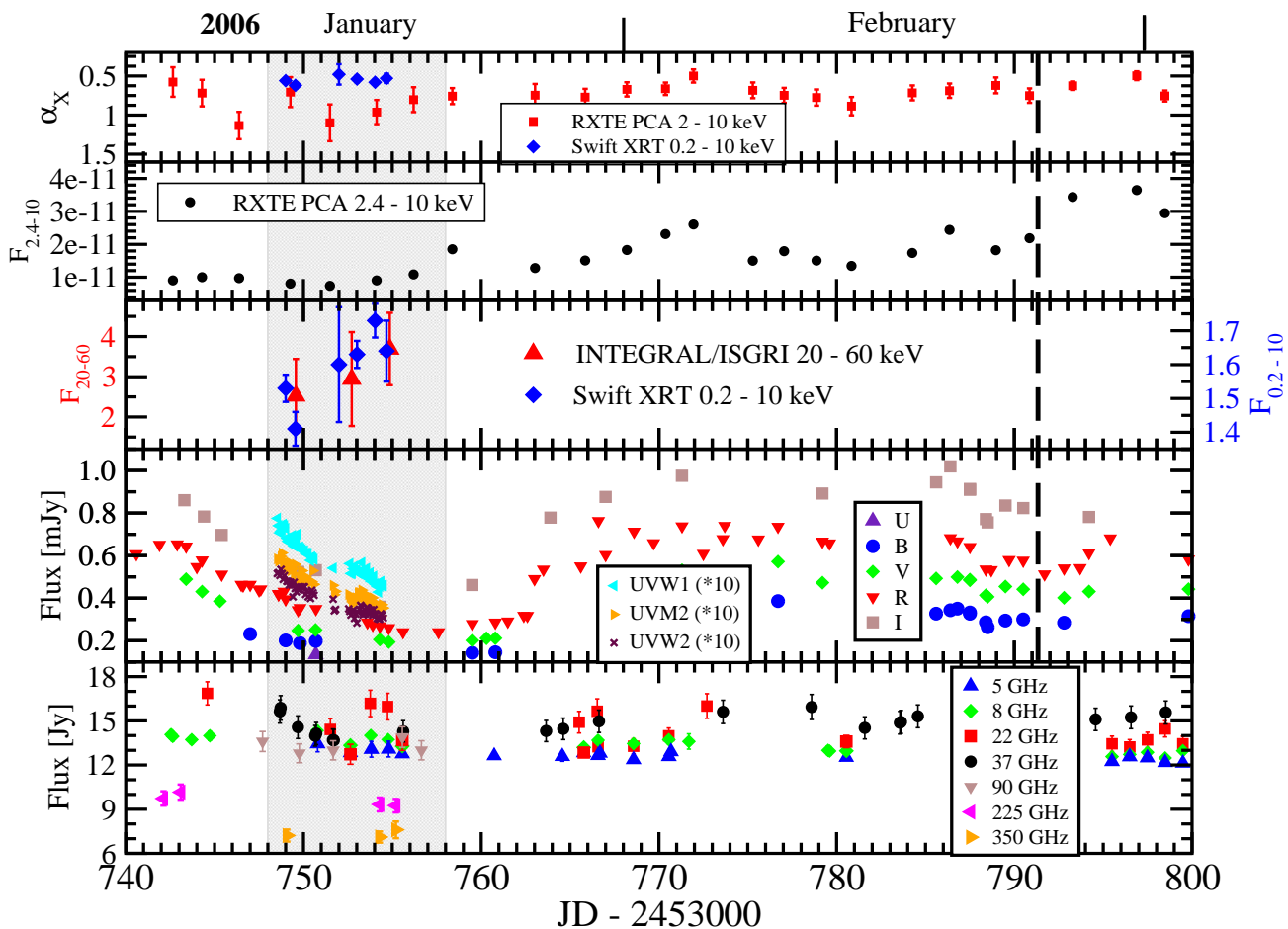}
\caption{Multiwavelength light curves of 3C~279 during January and
February 2006. The gray-shaded area indicates the core campaign period 
of the INTEGRAL, Chandra, and Swift observations; the vertical dashed 
line indicates the time of the MAGIC VHE $\gamma$-ray detection. Units
of the X-ray fluxes are: $F_{2-10}$ [erg~cm$^{-2}$~s$^{-1}$] and 
$F_{20-60}$ [$10^{-4}$~ph~cm$^{-2}$~s$^{-1}$]. The top panel shows 
the RXTE 2 -- 10~keV spectral index. The RXTE data are from \cite{Chatterjee08}, 
updated by A. Marscher (private comm.).
}
\label{mwlightcurve}
\end{figure*}

The quasar 3C~279 is regularly monitored at optical, near-infrared,
and radio frequencies by a number of observatories associated with
the Whole Earth Blazar Telescope (WEBT). It was the monitoring provided
by the WEBT optical partners that led to the discovery of the high state
of 3C~279 on 5 January 2006, which triggered our multiwavelength campaign.
Early results of this campaign were published and discussed by \cite{bbj07}. 
Here we add additional
observations in the �radio- to- sub-mm regime from radio observatories, which are
not asscociated with the WEBT, and which complement the spectral and time 
coverage considerably. The combined radio, sub-millimeter, optical, and X-ray data 
are shown in Figs.~ \ref{he_lc} -- \ref{mwlightcurve}.

In the optical, the WEBT monitoring of December 
2005 -- March 2006 indicated an overall
elevated optical state, with variability amplitudes of $\sim 0.5$~mag on
time scales of several days. The most remarkable light curve feature
during that campaign was an unusually clean, quasi-exponential decay of
the flux in all optical bands (BVRI) over a period of about two weeks 
($\sim$ JD 2453744 -- 2453757), including the core period 
illustrated in Figs.~\ref{he_lc} and \ref{mwlightcurve},
shortly after exceeding the trigger threshold of $R = 14.5$~mag. This decay
could be well approximated by an exponential $F(t) = F_0 \, e^{-t/\tau}$
with a time scale of $\tau \sim 12.8$~days, and continued throughout the 
time period of the high-energy (INTEGRAL, Chandra, Swift) observations. 
\cite{bp09} suggest that this clean, quasi-exponential decay may be the 
signature of deceleration of a ballistically moving plasmoid along the 
jet. 

The light curve recovered to a high level right after the period of the 
high-energy observations (around 22 January) and 3C~279 remained optically 
bright until the end of March 2006. This includes the time of the MAGIC 
detection (23 February 2006), during which the optical light curve did not 
show any remarkable variability features. While \cite{bbj07} found hints 
for spectral hysteresis and a hard lag within the optical bands in light 
curve segments later during the WEBT campaign, there was no evidence for 
optical spectral variability during the multiwavelength core campaign period 
(JD 2453748 -- 2453758, i.e., 12 -- 22 January 2006). 

\subsection{\label{Radio} Radio to sub-millimeter variability}
 The blazar 3C\,279 was observed in the radio- to sub-mm bands by a number of different observatories.
 The data sets are compiled from ongoing regular flux density monitoring programs
 and complemented by dedicated target-of-opportunity observations performed during
 2005 and 2007. The total of all flux density measurements cover a 
 frequency range from  1.4 to 345\,GHz, with the following contributions:
 The 100\,m Effelsberg telescope of the Max-Planck-Institut 
 f\"ur Radioastronomie (Bonn, Germany)
 observed at 1.4, 2.7, 5, 8.4, 10.5, 15, 22, 32 and 43\,GHz, providing flux density 
 measurements which are quasi-simultaneous in time (for details see:
 \cite{Fuhrmann08}, \cite{Angelakis08}). 

The IRAM 30\,m telescope on Pico Veleta (Spain) contributed with flux density measurements
at 90 and 230\,GHz. These data resulted from the regular IRAM flux monitoring program
(see \cite{Ungerechts98} and \cite{Agudo06} for a decription of the 
data reduction), and additional data obtained on the basis of target of opportunity observations 
proposed by us. These flux density measurements are complemented by additional 
measurements at 90 and 230 GHz with the 6x15\,m IRAM interferometer on Plateau
 de Bure (France) and at 90\,GHz with the Australia Telescope Compact Array (ATCA).
 To fill some gaps in the time coverage, we also included
 data from the VLA polarization monitoring, which is performed
 at 5 GHz, 8.4 GHz, 22 and 43\,GHz (\cite{Taylor00}). We further included
 the 14.5 GHz data from the University of Michigan Radio Observatory (UMRAO)
 monitoring program and the 230 and 345\,GHz data from the AGN monitoring of
 the Submillimeter Array (SMA) on Mauna Kea (Hawaii). The data from UMRAO and the SMA
 were already published in part by \cite{Chatterjee08} 
 and \cite{Larionov08}, where also the corresponding data reduction and calibration 
 is described. 
 
 In Fig.~\ref{radioplot1} we plot the flux density of 3C\,279 in the radio 
 bands versus time for the period 2004.4 - 2007.8. We have also added the
 optical R-band flux density from \cite{Chatterjee08}.
 In the lower panel of the figure we show 
 the flux-density variability of the VLBI core component 
 at 43\,GHz (same data as in \cite{Chatterjee08} but fitted
 with a circular Gaussian beam). Furthermore we add the times of zero 
 separation from the VLBI core for the jet components C22 ($2005.18 \pm 0.06$), 
 C23 ($2006.415 \pm 0.15$), and ($2006.894 \pm 0.11$) for C24 
 (see \cite{Chatterjee08}, \cite{Larionov08}).

 From Fig.~\ref{radioplot1} it is obvious that the variability amplitudes
 are more pronounced in the optical and mm-bands than in the cm-regime, which is
 consistent with a synchrotron self-absorbed radio jet. Between 2005.9 and 2007.0
 the optical to radio fluxes rose, indicative of an optical/radio flare, which
 peaked by the end of year 2007. During the rising phase of this flare, a first 
 local flux density maximum at mm-wavelengths, respectively a plateau in the 14.5\,GHz data
 was reached around 2006.4-2006.5, about half a year 
 before the absolute maximum of the
 flux densities were reached (peak: 2006.9 - 2007.0). This local maximum is clearly 
 seen in the SMA data at 345\,GHz (0.8\,mm) and 230\,GHz (1\,mm) 
 and also in the VLBI core flux at 43\,GHz (7\,mm). This indicates
 the superposition and blending of two flares separated in time by about 0.4-0.5\,yr. 
 It is tempting to associate the ejection of the VLBI components C23 and C24 with
 the observed double-peak structure between 2006 and 2007 in the total flux densities
 and in the VLBI core flux. After 2007.0, the optical flux 
 density decayed very fast, followed by successively more shallow decays 
 between sub-mm and cm-wavelengths.
 
 Between frequencies of 345\,GHz and 90\,GHz, the variability 
 is fast and the light curves are 
 partially undersampled in time. With the present time sampling, the fastest significant
 variations appear on time scales of $\leq 20 - 40$\,days. Faster variations cannot be
 excluded, but at this point remain speculative. At 90\,GHz a flux
 density increase from 12.8 in 2006.036 to 17.7 Jy in 2006.125 is observed in 34 days.
 This corresponds to a rate of 0.14 Jy/day for the flux density scale. This in turn leads to
 an apparent brightness temperature of the variable component 
 of $T_B \geq 8.4 \times 10^{12}$\,K,
 which could be reduced to values below the inverse-Compton limit of $10^{12}$\,K,
 if the Doppler-factor is larger than $\delta \geq 4.4$ (adopting the equations given
 in \cite{Fuhrmann08}). With this Doppler-factor the corresponding linear dimension
 of the emission region would be $ 42 / \delta = 9.5$\,$\mu$as, which corresponds
 to a linear scale of $r \leq 7.9 \times 10^{17}/ \delta= 1.8 \times 10^{17}$\,cm. 
 Using the above Doppler factor and a typical speed of the VLBI components of
 $\beta_{app} \simeq 15$\,c, we obtain a minimum bulk Lorentz-factor of the jet
 of $\gamma_{min} \geq 15$ and an inclination of the inner jet with respect to
 the line of sight at an angle of $i \leq 7^\circ$.

\subsection{Quasi-simultaneous radio spectra}

From the available radio data presented in the previous section, it is possible
to construct quasi-simultaneous broad-band radio spectra. Because of the irregular
time sampling at the different observatories, the binning of the data in time is different
for each epoch and was chosen to obtain the best possible frequency coverage. 
In Fig.~\ref{radioplot2} we show some examples of the radio spectra, 
which cover the frequency range from 
1.4 to 350\,GHz. The legend at the top left of the figure shows the observing date and
binning interval. The spectra are remarkably complex and show spectral evolution on
time scales of weeks to months. The spectrum of December 2005 (circles) was observed just
at the time before the 2006-2007 activity started and the 2007 flare began. 
The radio spectrum is flat up to about
43\,GHz, and steepens toward shorter (sub-mm) wavelengths ($\alpha_{43/225 GHz} \simeq 0.25$,
$\alpha_{225/345 GHz} \simeq 0.9$. From January 2006 to May 2006,
the spectrum becomes more and more inverted, with a prominent spectral maximum near 100\,GHz.
In the context of the three-stage shock in jet model of \cite{MarscherGear85} (see also
\cite{Lindfors06}), this behavior can be interpreted as a spectral component, which
is in its inverse-Compton stage (rising turnover flux $S_m$ and decreasing turnover frequency
$\nu_m$). Between 10 - 100 GHz, the observed spectral slopes vary considerably in all epochs. 
This suggests another variable spectral component with lower turnover
frequency, which is less prominent than the one
discussed before. In February 2006, this spectral component is best visible,
causing a spectral break near 43\,GHz, and it may have propagated down
to 15\,GHz by May 2006. Owing to the lack of multi-frequency VLBI data during this
time, it is not possible to decompose the radio spectrum of the total flux 
into several spectra of distinct VLBI jet components. However, with regard to the published
43\,GHz VLBI maps and the known jet kinematics (\cite{Chatterjee08}, \cite{Larionov08}),
it is very likely that the more prominent spectral break near 100\,GHz is associated
with a new jet component, which must have emerged from the self-absorbed VLBI core region 
between mid-January and mid-February 2006. A likely candidate for an
association of that spectral feature with a new jet component would be
C23, which became visible at 43\,GHz after $t_0 =2006.415 \pm 0.15$ 
(see solid black arrow in Fig.~\ref{radioplot2}). We note that an 
association of a secondary and less pronounced spectral
break in the 20-30 GHz range with an evolving older jet component (C22 
ejected at $\simeq 2005.2$) is a more speculative, but not unreasonable 
interpretation.

\section{\label{radioint}Interpretation of the radio data}

 The close correlation of the VLBI
 core flux with the total flux (see Fig.~\ref{radioplot1}, and \cite{Chatterjee08}, \cite{Larionov08})
 allows us to attribute most of the variability to be located within the synchrotron self-absorbed 
 VLBI core, for which an upper limit to its size $\theta$ is available  
 ($\delta \geq \theta D_L / (c t_{var}) /(1+z)$, with luminosity distance 
 $D_L$ and variability time scale $t_{var}$): $\theta \leq 0.03$\,mas or $r \leq 5.7 \times 10^{17}$\,cm.
 Applying the light travel time argument and following \cite{Jorstad05}, a lower
 limit to the relativistic Doppler factor of  $\delta \geq 5.1$ is obtained, which is close
 to the Doppler factors estimated in the previous sections. For synchrotron self-absorption the magnetic
 field can be calculated via 
 $ B_{ssa} [G] \propto 2.3 \times 10^{-5} \Delta S^{-2} \nu_m^{5} \theta^4 \delta /(1+z)$.
 This leads to $B_{ssa} \simeq 15$\,mG for the VLBI core and inner jet, 
 where we adopted $\nu_m=90$\,GHz for the spectral turnover frequency
 (see previous section and Fig.~\ref{radioplot2}). If we assume that most
 of the synchrotron energy is radiated near the turnover frequency $\nu_m$, we can estimate
 the Lorentz factor of the electrons via the relation $\nu_m = 1.2 \times 10^{-3} B \gamma^2$
 ($B$ in [G], $\nu_m$ in [GHz]) to be $\gamma \sim 2200$, which corresponds to an electron energy of 
 $E_e = \gamma m_e c^2 \simeq 1.1$\,GeV. First order inverse-Compton scattering
 would boost this radiation to $\nu_{ic} \propto \gamma^2 \nu_m \sim 10^{17.6}$\,Hz, thus
 into the X-ray domain. 
 This suggests that the X-ray emission of 3C~279 is dominated
 by synchrotron-self-Compton emission, while a different component
 (either Compton scattering of external photons or second-order
 Compton scattering of synchrotron photons) is needed to reproduce 
 the gamma-ray emission near $\sim 10^{24}$\,Hz, in agreement with 
 our SED modeling (see Sect.~\ref{modeling}).

\section{\label{SED}Spectral energy distribution}

\begin{figure}[ht]
\vskip 1cm
\includegraphics[width=\columnwidth]{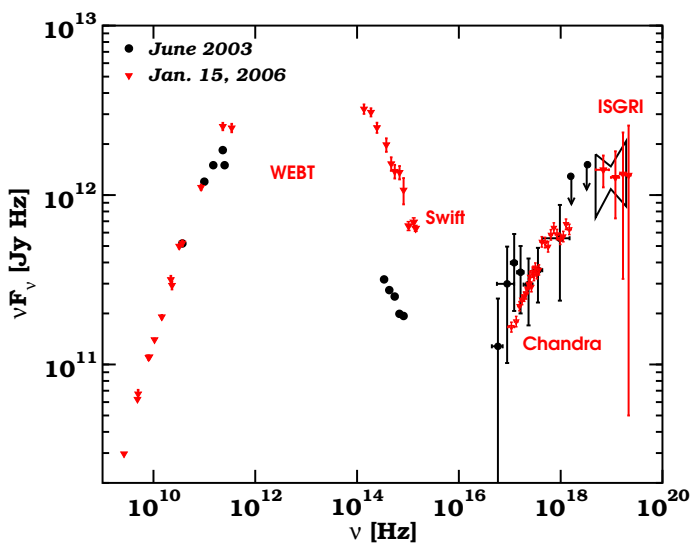}
\caption{Spectral energy distribution of 3C~279, centered on 15 January 2006
(red downward triangles). All IR/optical/UV data were taken within $\pm 1/2$~hr 
of UT 05:00. Also shown for comparison is the SED from the INTEGRAL AO-1 campaign 
from June 2003 (\cite{Collmar04}). Despite the huge difference in optical flux, the 
X- and hard X-ray fluxes are surprisingly similar.   
}
\label{SED0}
\end{figure}

The best multiwavelength coverage at IR/optical/UV frequencies was obtained 
on 15 January 2006, where observations in the near-IR, optical (UBVRI) and
UV (UVW1, UVM2, UVW2) frequencies were available within $\sim 1/2$~hr of 
UT 05:00. We therefore focus our attention on the simultaneous multiwavelength 
SED centered on that time. For the X-ray regime, we included the high-quality
Chandra spectrum of 17 January 2006 in our SED, which is justified
by the moderate X-ray variability during the core campaign. 
Given the substantial integration time required to obtain spectral information 
from the INTEGRAL observations, we use the time-averaged spectrum over the entire 
observation for inclusion in the SED. Justified
by the long variability time scales in the radio, data taken within a few 
days of 15 January 2006 were included to extend the SED into the radio 
regime. The resulting SED, including radio, IR, optical (WEBT), UV (Swift UVOT)
and X-ray (Chandra, INTEGRAL) and soft $\gamma$-ray (INTEGRAL) 
is displayed in Fig.~\ref{SED0} and compared to
the very-low-state SED from the multiwavelength campaign during INTEGRAL AO-1 
(June 2003, \cite{Collmar04}). 

The most remarkable feature of the 15 January 2006 SED is the low X-ray 
flux, while the optical flux was clearly in an elevated state. The X-ray
spectrum is virtually indistinguishable from the historical low-state SED
of June 2003, while the optical flux is about a factor of 10 higher. 
The only slight difference in the X-ray spectral shape is the indication
of a steepening toward lower frequencies, possibly indicating a low-energy
cutoff of the X-ray spectrum (and hence, the underlying electron distribution,
see Sect.~\ref{modeling}), for which there was no evidence in the June 2003
SED. A comparison to a further historical SED is shown in Fig.~\ref{SEDfits}
\footnote{\tt The interested reader may contact W. Collmar for
an ASCII table of the figure.}.

Unfortunately, because the time of this campaign was before the launch of
AGILE and Fermi, there was no simultaneous coverage of the critical MeV
-- GeV $\gamma$-ray regime, where the high-energy component of 3C~279 is
expected to peak. Therefore, constraints on the peak of the high-energy
component can only be based on plausibility arguments in comparison with
historical SEDs, an example of which is shown in Fig.~\ref{SEDfits}.

\section{\label{modeling}SED modeling}

As demonstrated in \cite{Hartman01a}, a leptonic one-zone jet model, including
both SSC and external-Compton scattering as high-energy radiation mechanisms, 
has been appropriate and very successful for modeling all simultaneous SEDs
of 3C~279 collected during the lifetime of EGRET. We therefore choose the same
approach to interpret the low X-ray flux states presented in Fig.~\ref{SED0},
including the SED from 15 January 2006, the center of the core period of our
multiwavelength campaign described here. For this purpose, we use the equilibrium
version of the leptonic SSC + EC model of \cite{bc02}. 

In this model, a relativistic electron population is assumed to be in an
equilibrium state between on-going injection, escape, and radiative cooling in 
a spherical emission region of radius $R$, moving relativistically with bulk 
Lorentz factor $\Gamma$ along the jet. Electrons are injected with a power-law
distribution specified by an injection spectral index $q$ and low- and high-energy
cutoffs $\gamma_1$ and $\gamma_2$, respectively. The radiative cooling is 
evaluated self-consistently with the radiative output via synchrotron, SSC, 
and external-Compton emission. 
Particles are assumed to escape on an energy-independent time
scale $t_{\rm esc} \equiv \eta R/c$, where $\eta > 1$ is a free
parameter. Particle escape will become relevant for particles 
with energies below a critical break energy $\gamma_b$ defined
through the condition that $t_{\rm esc} = t_{\rm cool}$, the
radiative cooling time scale. Escaping particles are leaving 
the emission region in random directions and will therefore 
primarily end up in a slower sheath that may be present around
the fast spine producing the high-energy emission.
The magnetic field $B$ is a free parameter
of the model, as are the parameters characterizing the external radiation
field. For any given set of parameters, the code also evaluates the ratios
of co-moving energy densities in the magnetic field, $u'_B$ and the equilibrium
electron distribution, $u'_e$, the corresponding kinetic ($L_e$) and Poynting 
flux ($L_B$) powers, and their ratio, $\epsilon_B \equiv u'_B / u'_e = L_B / L_e$. 
For a more detailed description of this equilibrium model see \cite{acciari09}. 

\begin{figure*}[ht]
\vskip 1cm
\includegraphics[width=17cm]{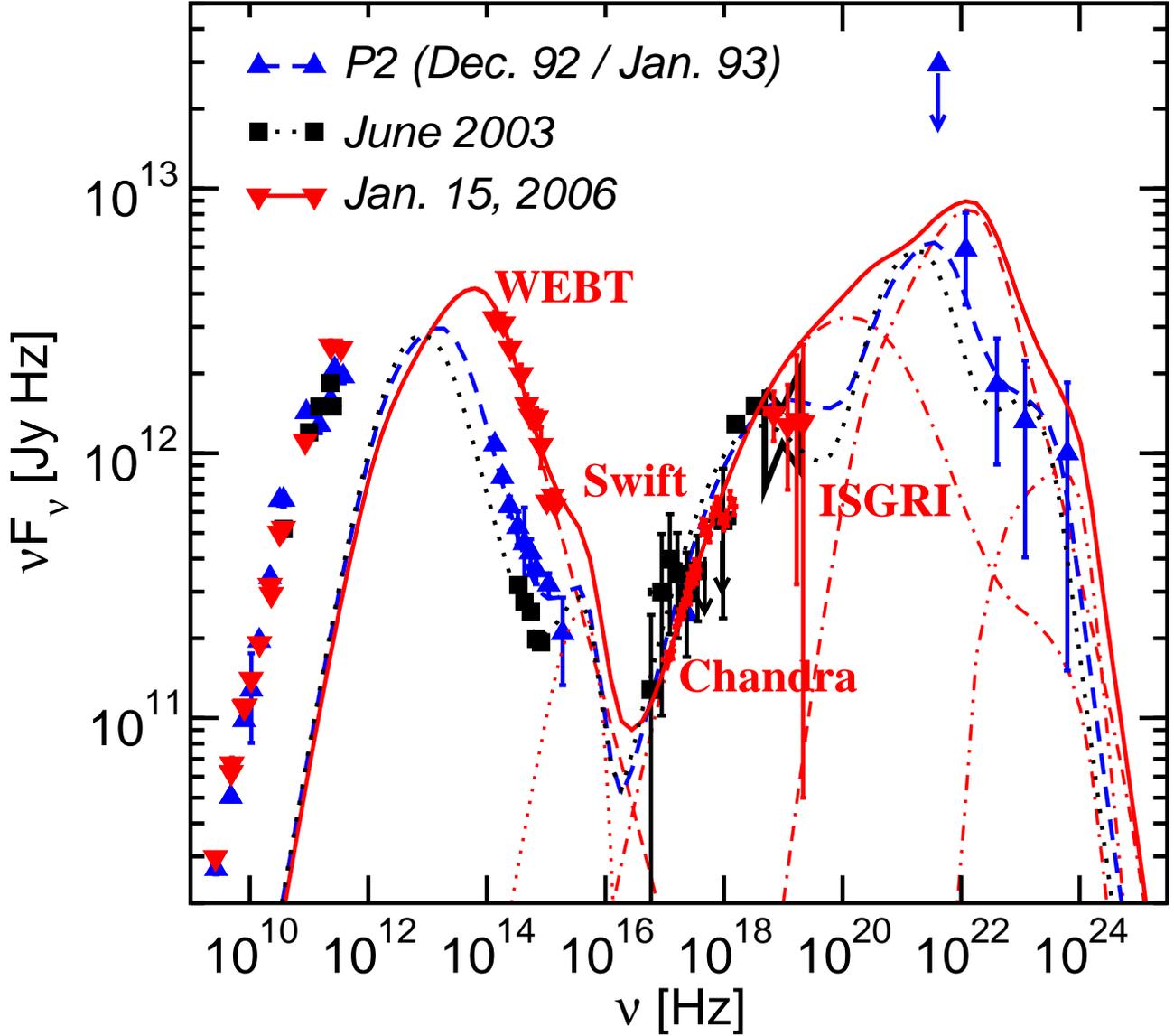}
\caption{Fits to the two low-X-ray state SEDs included in Fig.~\ref{SED0}, plus 
a historical low-X-ray state SED from the CGRO era in December 1992 / January 1993.
The individual model radiation components for the 15 January 2006 SED are shown.   
See text for details.}
\label{SEDfits}
\end{figure*}

The model fits only the near-infrared to gamma-ray emission, which is believed to be produced 
at the sub-pc scales of the jet, where the source is still optically thick at 
radio wavelengths. 
The optically thick radio emission is believed to be produced on $>$ pc scales.
Above $\sim 100$\,GHz however, the emission comes from sub-parsec scales of
a region which is located at or very near the base of the VLBI jet (size of
VLBI core at 100 GHz $\leq 10^{18}$\,cm).
We find that the optical through X-ray SED of 15 January 2006 can be well
represented with the following model parameters: $L_e = 1.2 \times 
10^{45}$~ergs~s$^{-1}$, $\gamma_1 = 1.5 \times 10^3$, $\gamma_2 = 10^5$,
$q = 3.7$, $B = 0.8$~G, $R = 3.5 \times 10^{16}$~cm, $\Gamma = 15$, 
and an external radiation field energy density $u_{\rm ext} = 2.2 \times 
10^{-4}$~ergs~cm$^{-3}$. In order to reduce the number of free parameters,
we choose the observing angle in a way that the Doppler factor $D \equiv
\left( \Gamma [ 1 - \beta_{\Gamma} \cos\theta_{\rm obs} ] \right)^{-1}
= \Gamma$. The magnetic field specified above corresponds to a value
slightly below equipartition with $\epsilon_B = 0.56$. The fit is illustrated
by the red solid curve in Fig.~\ref{SEDfits}.
While the X-ray spectrum is always produced through 
SSC emission, the $\gamma$-ray emission is strongly dominated 
by the EC emission.
The model does, indeed, slightly over-predict the soft
$\gamma$-ray (ISGRI) flux. However, the ISGRI error bars are
large, and our fit can still be considered marginally 
consistent with the ISGRI spectrum.

Remarkably, the other two low-X-ray states can be modeled with essentially
identical parameters, changing only the low-energy cutoff of the electron
distribution. The fits to the P2 and June 2003 spectra shown in 
Fig.~\ref{SEDfits} have been generated choosing $\gamma_1 = 750$ for P2 and
$\gamma_1 = 550$ for June 2003. As already alluded to before, the higher
value of the low-energy cutoff may be responsible for the break in the
X-ray spectrum seen in the Chandra data of our January 2006 campaign.

The change in $\gamma_1$ affects the peak frequencies of all 
 radiation components as $\nu_{\rm sy} \propto \gamma_1^2$,
 $\nu_{\rm EC} \propto \gamma_1^2$, and $\nu_{\rm SSC} \propto
 \gamma_1^4$, as long as Compton scattering occurs in the Thomson
 regime near the Compton peaks, which is the case for the parameters
 adopted here. In addition, increasing $\gamma_1$ also has the
 effect of increasing the synchrotron radiation energy density,
 hence leading to an increasing SSC contribution in the high-energy
 SED. In Fig.~\ref{SEDfits} we included the individual radiation
 components - synchrotron, SSC, EC (BLR), EC (disk), and the 
 disk emission as dashed, dot-dashed, double-dot-dashed, 
 double-dash-dotted and dotted curves for the 15 January 2006 
 SED fit. This illustrates the dominant SSC contribution 
 to the soft $\gamma$-ray regime in this fit. In the states 
 fitted with lower values of $\gamma_1$, the SSC component still 
 dominates in the X-ray band, but plays virtually no role in the 
 production of the $\gamma$-ray emission.

In the model calculations for 3C~279, all our cases are in the
fast-cooling regime. In the fast-cooling regime, particles will
cool below $\gamma_1$ to a cutoff energy determined by a
balance between radiative cooling and escape. In an internal
shock scenario, the value of $\gamma_1$ may be related to the
relative Lorentz factor of two colliding shells, and the efficiency
of transferring swept-up proton energy to accelerated electrons.
Because we do not fully understand the details of this transfer
process, this latter effect is commonly parameterized through
electron acceleration parameters $\epsilon_e$ and $\zeta_e$,
both assumed to be on the order of $\sim 10^{-1}$ (e.g., \cite{bd10}).
Those parameters are defined so that $\epsilon_e$ 
is the fraction of swept-up proton energy
transferred to relativistic electrons, and $\zeta_e$ 
the fraction of swept-up electrons that are accelerated to
relativistic energies. Specifically, if the electron acceleration
process results in an electron spectral index of $q > 2$, the
low-energy cutoff is given by
        
\begin{equation}
\gamma_1 = {m_p \over m_e} {q - 2 \over q - 1} \, {\epsilon_e
\over \zeta_e} \, (\Gamma_{\rm shock} - 1) ,
\end{equation}
where $\Gamma_{\rm shock}$ is the Lorentz factor of the (internal)
forward or reverse shock resulting from the collision of two
shells of relativistic ejecta, measured in the co-moving frame
of the shocked material (\cite{bd10}). While
$\Gamma_{\rm shock}$ is a strong function of the relative
Lorentz factors of the colliding shells, $\epsilon_e$ and
$\zeta_e$ are expected to depend on the efficiency of the
generation of turbulent magnetic fields mediating the energy
transfer between protons and electrons behind the shock fronts.
 
In \cite{bbj07}, constraints on jet parameters
could be derived from the estimated synchrotron peak flux
in the 15 January 2006 SED, and a hint for hard
lags among the optical (BVR) bands. The synchrotron peak
flux was translated into a magnetic-field estimate according
to Eq.~4 in \cite{bbj07}. Inserting the
values of our fit results, $D = 15$ and $\epsilon_B = 0.56$,
that estimate becomes

\begin{equation}
B \ge B_{\epsilon_B} = 0.74 \left( t_{\rm var, d}^{\rm obs} 
\right)^{-1} \; {\rm G} ,
\end{equation}
which perfectly agrees with the adopted value $B = 0.8$~G in
our fit. 

Based on evidence for hard lags (R band leading the V and B
band fluxes), \cite{bbj07} developed even more
parameter constraints based on the assumption that the lags
are caused by a slow acceleration process, which needs to be
faster than the radiative cooling time scales. This led to
the magnetic-field estimate in Eq.~6 in \cite{bbj07}. 
Using our fit values as well as the best-fit B - R
time lag, $\tau_{BR} = 3.75$~d, we find
  
\begin{equation}
B \le B_{\rm acc} = 0.20 \, (1 + k)^{-2/3} \; {\rm G} ,
\end{equation}
where $k$ is the ratio of Compton-to-synchrotron cooling 
rates (i.e., $k = u'_{\rm rad} / u'_B$ if Compton scattering
occurs in the Thomson regime). This upper limit is substantially
lower than our adopted fit value. However, we point out 
that evidence for hard lags in the WEBT campaign data seems 
to have been most prominent in the second half of the campaign, 
where even hints for spectral hysteresis among the optical 
bands were found. No indication of spectral hysteresis was 
present during the time corresponding to the 15 January SED 
fitted in this paper.

\section{Summary and conclusions}

We have observed the prominent FSRQ 3C~279 over a period of about 10 days
in an intensive multiwavelength campaign in January 2006. We obtained hard
X-ray / soft $\gamma$-ray coverage with INTEGRAL, additional X-ray monitoring
observations with Swift XRT, and RXTE, a high-quality 0.3 -- 7~keV X-ray 
spectrum from Chandra, optical/UV monitoring with Swift/UVOT, and also
long-term optical/near-IR/radio monitoring by the WEBT, and the Effelsberg -,  
UMRAO, IRAM, and ATCA radio telescopes, and the SMA.

Our campaign caught the object initially in a high optical state, followed
by a clean, quasi-exponential decline throughout the core campaign period.
The X-ray monitoring observations indicate the opposite trend, namely a
gradual flux increase. The optical and X-ray fluxes of 3C~279 clearly 
do not follow a correlated variability trend during our campaign. 

The campaign results allowed us to construct a well sampled SED from radio
through soft $\gamma$-rays. Comparing this SED to a historical SED from the
EGRET era as well as INTEGRAL AO-1 (June 2003), we find that among EGRET
P2 (December 1992 -- January 1993), INTEGRAL AO-1 (June 2003) and our 
January 2006 campaign, the optical flux level --- but not the optical
continuum spectral shape --- had changed by almost one order of magnitude,
while the X-ray flux and spectral shape appear almost unchanged. Our 
January 2006 SED represents the highest optical flux state among the three
SEDs compared here, and is the only SED in which we found evidence for
a downward bending in the X-ray spectrum, possibly indicating a low-energy
cut-off. 

We modeled the SED of January 2006 together with the other two low-X-ray 
flux SEDs mentioned above, with a one-zone leptonic model, including synchrotron,
SSC, and external-Compton emission and self-consistently evaluating an equilibrium
electron distribution with the radiative output. This model was previously
successfully applied to all simultaneous SEDs of 3C~279 collected during the
CGRO era. We find that all three low-X-ray SEDs can be well represented with 
this model. Most remarkably, the optical variability among these three SEDs
can be reproduced by changing only the low-energy cutoff of the electron
distribution, leaving all other parameters fixed. We caution that this 
interpretation of the variability is obviously not unique, given the substantial
number of parameters entering the model. However, the evidence for a low-energy 
break in the January 2006 X-ray spectrum lends additional support to the presence 
of a low-energy cutoff in the electron spectrum at ultrarelativistic energies. 
This result may have important consequences for the evaluation of the total
kinetic power carried by the jet in 3C~279. Given the steep spectral index
of the electron spectrum inferred from the steep optical spectrum, most of
the kinetic energy in the leptonic component of the sub-pc scale jet is 
carried by the low-energy end of the electron distribution. Because we lack evidence
(e.g., a bulk Compton component, \cite{bs87}) for a substantial thermal 
plasma component in the jet, our results then suggest that the kinetic luminosity 
in relativistic electrons required for our model fit ($L_e = 1.2 \times 
10^{45}$~erg~s$^{-1}$) provides a realistic estimate of the total kinetic 
luminosity in leptons in the relativistic jet.

\begin{acknowledgements}
We thank S. Jorstad and A. Marscher for providing VLBI data at 43 GHz
and for discussion. We thank the IRAM Granada staff for data from their 
regular flux monitoring program, and in particular H. Ungerechts and S. Leon 
for their support.
This research is based on observations with the Effelsberg 100-m telescope of the 
Max-Planck-Institut f\"ur Radioastronnomie (Bonn, Germany). This research has made
use of data from the University of Michigan Radio Astronomy Observatory (UMRAO), which
has been supported by the University of Michigan and the National Science Foundation.
This work has made use of observations with the IRAM-interferometer and IRAM 30-m telescope,
the Australia Telescope Compact Array, the Very Large Array (NRAO) and the
Sub-millimeter Array (SMA).
The Submillimeter Array is a joint project between the Smithsonian 
Astrophysical Observatory and the Academia Sinica Institute of Astronomy and 
Astrophysics and is funded by the Smithsonian Institution and the Academia 
Sinica. M.B. acknowledges support from NASA through INTEGRAL Guest Investigator
grant NNX09AI71G.
{\it Swift} is supported at Penn State by NASA contract NAS5-00136.
D.G. acknowledges support by NASA contract NNX07AH67G.
I.A. acknowledges support by the National Science Foundation of the USA and 
the ``Ministerio de Ciencia e Innovaci\'on" of Spain through grants 
AST-0907893, and AYA2007-67627-C03-03, respectively.

\end{acknowledgements}

\end{document}